\documentclass[aps,prd,reprint,preprintnumbers,superscriptaddress,showpacs,twocolumn]{revtex4-1}
\usepackage{latexsym,graphicx,amssymb,amsmath,mathrsfs}
\usepackage{setspace,bm}
\usepackage[breaklinks, colorlinks=true, pdfstartview=FitV, linkcolor=red, citecolor=blue, urlcolor=blue]{hyperref}
\usepackage[usenames]{color}
\usepackage{latexsym}
\usepackage{epstopdf}
\usepackage{mathtools}
\usepackage{amssymb}
\usepackage{fancybox}

\usepackage[utf8]{inputenc}
\usepackage{tikz}
\usetikzlibrary{shapes,arrows}

\usepackage{CJKutf8}

\allowdisplaybreaks[1]
\usepackage[normalem]{ulem}  

\renewcommand\sout[1]{\bgroup \color{red} \ULdepth=-.5ex \ULset {#1}}

\begin{document}
\preprint{YITP-22-97}

\title{
Improving efficiency of the path optimization method for a gauge theory}

\author{Yusuke Namekawa}
\email[]{namekawa@hiroshima-u.ac.jp}
\affiliation{Education and Research Center for Artificial Intelligence and Data Innovation, Hiroshima University, Hiroshima 730-0053, Japan}
\affiliation{Department of Physics, Faculty of Science, Kyoto University, Kyoto 606-8502, Japan}
\affiliation{Yukawa Institute for Theoretical Physics, Kyoto University, Kyoto 606-8502, Japan}

\author{Kouji Kashiwa}
\affiliation{Department of Computer Science and Engineering, Faculty of Information Engineering, Fukuoka Institute of Technology, Fukuoka 811-0295, Japan}

\author{Hidefumi Matsuda}
\affiliation{Department of Physics, Faculty of Science, Kyoto University, Kyoto 606-8502, Japan}
\affiliation{Department of Physics and Center for Field Theory and Particle Physics, Fudan University, Shanghai 200433, China}

\author{Akira Ohnishi}
\affiliation{Yukawa Institute for Theoretical Physics, Kyoto University, Kyoto 606-8502, Japan}

\author{Hayato Takase}
\noaffiliation{}

\begin{abstract}
We investigate efficiency of a gauge-covariant neural network and an approximation of the Jacobian in optimizing the complexified integration path toward evading the sign problem in lattice field theories.
For the construction of the complexified integration path, we employ the path optimization method.
The $2$-dimensional $\text{U}(1)$ gauge theory with the complex gauge coupling constant is used as a laboratory to evaluate the efficiency.
It is found that the gauge-covariant neural network, which is composed of the Stout-like smearing, can enhance the average phase factor, as the gauge-invariant input does.
For the approximation of the Jacobian, we test the most drastic case in which we perfectly drop the Jacobian during the learning process.
It reduces the numerical cost of the Jacobian calculation from ${\cal O}(N^3)$ to ${\cal O}(1)$, where $N$ means the number of degrees of freedom of the theory.
The path optimization using this Jacobian approximation still enhances the average phase factor at expense of a slight increase of the statistical error.
\end{abstract}

\maketitle

\section{Introduction}

The Monte-Carlo (MC) method is an important tool for investigating non-perturbative properties of quantum field theories in which we do not know the analytic results.
The MC method with the probability weight is, however, difficult to be performed in high precision when the sign problem arises.
The probability distribution function that controls the update of the configuration becomes complex and/or seriously oscillating functions.
Actually, quantum chromodynamics (QCD) at finite density is a well-known theory that has a serious sign problem: for example, see~\cite{Nagata:2021ugx}.
Recently, several new methods have been developed to control the sign problem at high densities.
One of the famous examples is the complex Langevin method~\cite{Klauder:1983sp,Parisi:1984cs}, which is based on a stochastic quantization with complexified dynamical variables and is free from the sign problem.
Thanks to the low numerical cost of the complex Langevin method, it has already been applied to 4-dimensional QCD at finite density~\cite{Sexty:2013ica,Aarts:2014bwa,Fodor:2015doa,Nagata:2018mkb,Kogut:2019qmi,Sexty:2019vqx,Scherzer:2020kiu,Ito:2020mys}.
However, the validity region of the complex Langevin method is limited.
The complex Langevin method is proved to provide correct results if the boundary term disappears~\cite{Aarts:2009uq}, or equivalently, the distribution of the driving force decays exponentially or faster~\cite{Nagata:2016vkn}.
Another famous method is the tensor renormalization group method~\cite{Levin:2006jai}, which is based on a coarse graining non-MC algorithm using a tensor network.
Although the computational cost is extremely high, it has been vigorously tested even in 4-dimensional theoretical models~\cite{Akiyama:2019xzy,Akiyama:2020ntf,Akiyama:2020soe,Akiyama:2021zhf} as well as non-Abelian theories~\cite{Bazavov:2019qih,Asaduzzaman:2019mtx,Fukuma:2021cni,Hirasawa:2021qvh,Kuwahara:2022ubg}.
The improved algorithms are proposed to reduce the enormous computational cost~\cite{Kadoh:2019kqk,Kadoh:2021fri}.
The Lefschetz thimble method~\cite{Witten:2010cx} is a MC scheme that complexifies dynamical variables and determines the integration path by solving an anti-holomorphic flow equation from fixed points such that the imaginary part of the action is constant.
Cauchy's integral theorem ensures that the integral is independent of a choice of the integration path if the path is given as a result of continuous deformation from the original path~\cite{Alexandru:2015sua}, crosses no poles of the integrand, and the integral at infinity has no contribution.
The numerical study has been started with Langevin algorithm~\cite{Cristoforetti:2012su}, Metropolis algorithm~\cite{Mukherjee:2013aga}, and Hybrid MC algorithm~\cite{Fujii:2013sra}.
The additional ergodicity problem and the high computational cost are the main bottlenecks of this method, but the algorithm development is overcoming them~\cite{Fukuma:2020fez,Fukuma:2021aoo}.
The path optimization method (POM)~\cite{Mori:2017pne,Mori:2017nwj}, also known as the sign-optimized manifold~\cite{Alexandru:2018fqp,Lawrence:2022afv}, is an alternative approach that modifies the integration path using machine learning through neural networks.
Machine learning finds the best path on which the sign problem is maximally weakened.
The POM successfully works for the complex $\lambda \phi^4$ theory~\cite{Mori:2017pne}, the Polyakov-loop extended Nambu--Jona-Lasinio model~\cite{Kashiwa:2018vxr,Kashiwa:2019lkv}, the Thirring model~\cite{Alexandru:2018fqp,Alexandru:2018ddf}, the $0+1$ dimensional Bose gas~\cite{Bursa:2018ykf}, the $0+1$ dimensional QCD~\cite{Mori:2019tux}, the 2-dimensional $\text{U}(1)$ gauge theory with complexified coupling constant~\cite{Kashiwa:2020brj}, the 2+1-dimensional XY model~\cite{Giordano:2022miv}, as well as noise reduction in observables~\cite{Detmold:2021ulb}.
The recent progress of the complexified path approaches is reviewed in Ref.~\cite{Alexandru:2020wrj}.

A crucial issue of the POM in gauge theories is control of redundancy from the gauge symmetry.
In $0+1$ dimensional QCD at finite density~\cite{Mori:2019tux}, the POM works with and without the gauge fixing.
In higher dimensions, however, the gauge-fixing or gauge-invariant input is required for neural networks to find an improved integral path.
The 2-dimensional $\text{U}(1)$ gauge theory with a complex coupling~\cite{Kashiwa:2020brj,Pawlowski:2021bbu,Namekawa:2021nzu} is a good test ground to investigate the effect of gauge degrees of freedom on the sign problem.
The sign problem originates from the imaginary part of the complex coupling.
Since the analytic result is available~\cite{Wiese:1988qz,Rusakov:1990rs,Bonati:2019ylr}, we can use it to verify the simulation results.
It has been found that the average phase factor, an indicator of the sign problem, is not improved without the gauge fixing.
This may be related to the insufficient performance of the considering neural network in Ref.~\cite{Kashiwa:2020brj} compared to the complexity of the gauge symmetry. 
In contrast, the gauge-invariant input successfully enhances the average phase factor without the gauge fixing~\cite{Namekawa:2021nzu}.
The link variables are no longer a direct input to the neural network. Gauge-invariant quantities, plaquettes in this case, are chosen as the input.
This treatment mitigates the difficulties induced by the gauge symmetry, even in a simple neural network.
A similar idea is employed as a part of lattice gauge equivariant Convolutional Neural Networks~\cite{Favoni:2020reg}.

To tackle the sign problem in more realistic systems, it is hoped to develop efficient optimization methods further.
In this study, we try to implement the feasible ways to the path optimization method as summarized below.
\begin{description}
   \item[Improvement 1] gauge-covariant neural network
   \item[Improvement 2] Approximation of the Jacobian calculation in the learning process
\end{description}
The gauge-covariant neural network is an alternative neural network which respects the gauge symmetry~\cite{Tomiya:2021ywc}.
In the method, the usual neural network is replaced by the Stout-like smearing functions.
Since the smearing process does not break the gauge covariance by definition, we can exactly deal with the gauge symmetry of theories.
It is important to evaluate the performance of the gauge-covariant neural network toward large scale simulations.
It is also necessary to reduce the numerical cost to calculate the Jacobian.
The cost is $\mathcal{O}(N^3)$ with $N$ being the degrees of freedom.
We test the most drastic approximation.
We completely neglect the Jacobian in the learning part.
The Jacobian is calculated only at the configuration generation and the final step of the POM for measurement.

This paper is organized as follows.
In Sec.~\ref{sec:POM}, the current status of the path optimization method is summarized.
Improvements in the path optimization method using the gauge-covariant neural network and the approximation of Jacobian are discussed in Sec.~\ref{sec:improve}.
Section~\ref{sec:numerical_result} shows numerical results and Sec.~\ref{Sec:Summary} summarizes this paper.
In Appendix~\ref{sec:link}, we discuss improvements in the optimization using the link-variable input.

\section{Path optimization method}
\label{sec:POM}

We compile the current status of the path optimization method.
In the path optimization method, dynamical variables are complexified to control the sign problem.
The path-integral contour is deformed on the complexified dynamical variable plane.
Because of the Cauchy's integral theorem, such deformation provides the same result obtained on the original integral contour, as long as the modified path does not cross any poles of the integrand, the Boltzmann weight, and its infinities do not contribute to the result. Note that the singular points of the action are generally irrelevant. They appear from the zeros of the fermion determinant and result in the zeros, not poles, of the integrand.

The procedure of the path optimization method can be summarized as follows:
\begin{description}
   \item[Step 1] Create configurations on the original integral path using the MC method.
   \item[Step 2] Using the back-propagation method to optimize the network parameters to construct the modified integral path.
   \item[Step 3] Create configurations on the modified integral path using the reweighting method.
   \item[Step 4] Evaluate the average phase factor.
   \item[Step 5] Return to Step 2 if the growth of the average phase factor is not sufficient.
\end{description}
In addition, we occasionally need the replica exchange MC method~\cite{Swendsen:1986replica,Geyer:1991markov,Marinari:1992qd,Hukushima:1996}, as needed in the Lefschetz thimble method~\cite{Fukuma:2017fjq,Fukuma:2019wbv,Fukuma:2019uot}, when the integral path becomes complicated: for example, see Ref.~\cite{Kashiwa:2020brj}.
Details of the path optimization method are explained below.

\subsection{Gauge variant input}
The simplest way is to directly complexify the dynamical variables $A_\mu(n)$ in the path optimization method as
\begin{align}
A_\mu(n) \to {\cal A}_\mu(n) = A_\mu(n) + i z_\mu(n),
\end{align}
where ${\cal A}_\mu(n) \in \mathbb{C}$ and $A_\mu(n), z_\mu(n) \in \mathbb{R}$ at the lattice site $n$ in the direction of $\mu$.
Complexified variables, ${\cal A}_\mu(n)$, represent the modification of the integral path.
To construct the best $z_\mu(n)$ for reduction of the sign problem, we use the neural network, which obeys the universal approximation theorem~\cite{Cybenko:1989,Hornik:1991251}.
The actual procedure then becomes
\begin{align}
\underbrace{A_\mu(n)}_{\mathrm{input\,layer}} \to {\mathrm{hidden\,layer}} \to \underbrace{{\cal A}_\mu(n)}_\mathrm{output\, layer}.
\end{align}

In the lattice simulation, it is convenient to transform the above dynamical quantities into the link variables at the input layer because the lattice action is composed of the link variables:
\begin{align}
\underbrace{A_\mu(n) \to U_\mu(n)}_{\mathrm{input\,layer}} \to{\mathrm{hidden\,layer}} \to \underbrace{{\cal U}_\mu(n)}_\mathrm{output\, layer},
\end{align}
where $U_\mu(n) := \exp(i g_0 A_\mu(n))$ and ${\cal U}_\mu(n) := \exp(i g_0 {\cal A}_\mu(n))$ are the original and modified (complexified) link variables, respectively, with the bare coupling constant $g_0$.
Such compact variables make the path integral finite.

In the neural network with the input represented as $t_i$, corresponding to the real and imaginary parts of $U_\mu(n)$, the variables on the hidden layer ($y_j$) and the output ($z_k$) are given as
\begin{align}
    & y_j = F(w^{(1)}_{ji} t_i + b_j^{(1)}),
    \nonumber\\
    & z_k  = \omega_k F(w^{(2)}_{kj} y_j + b_k^{(2)} ),
    \label{Eq:FNN}
\end{align}
where $i = 1, ..., N_{\rm input}$, $j = 1, ..., N_{\rm hidden}$, and $k = 1, ..., N_{\rm output}$ with the numbers of units in the input, hidden, and output layers, $N_{\rm input}$, $N_{\rm hidden}$ and $N_{\rm output}$.
In the case of the gauge variant input, $N_{\rm input} = N_{\rm output} = 4 N_{\rm vol}$, where $N_{\rm vol} = N_1 N_2$ with $N_\mu$ being the lattice size in the $\mu$-direction.
$N_{\rm hidden}$ is taken to be proportional to $N_{\rm vol}$.
$w$, $b$ and $\omega$ are parameters of the neural network and $F$ is the so-called activation function.
We employ a hyperbolic tangent function for the activation function.

\subsection{Gauge invariant input}

For the learning process in the path optimization method, the gauge invariance plays a central role and thus should be imposed on the neural network.
One possible way is using the gauge invariant input.
The link variables are transformed into the plaquette on the input layer, which is the gauge-invariant quantity and the fundamental building block of the lattice Lagrangian density:
\begin{align}
&\underbrace{A_\mu(n) \to U_\mu(n) \to P_{\mu\nu}(n)}_{\mathrm{input\,layer}} \nonumber\\
&\hspace{3cm} \to{\mathrm{hidden\,layer}} \to \underbrace{{\cal U}_\mu(n)}_\mathrm{output\, layer},
\end{align}
where the plaquette $P_{\mu\nu}$ is defined as
\begin{align}
  P_{\mu\nu}(n) := U_{\mu}(n) \, U_\nu (n + \hat{\mu}) \, U^{-1}_\mu (n + \hat{\nu}) \, U_\nu^{-1}(n),
\end{align}
where $\hat{\mu}$ ($\hat{\nu}$) is a unit vector in the $\mu$ ($\nu$) direction.
In this case, $N_{\rm input} = 2 N_{\rm vol}$.

It is applicable to any lattice gauge action including the plaquette.
The path-optimized lattice action is represented by the modified plaquettes.
With this procedure, the hidden layer has the gauge invariance by definition.
This procedure corresponds to the data pre-processing known in the machine learning community.
Its effectiveness is confirmed in the 2-dimensional $\text{U}(1)$ gauge theory~\cite{Namekawa:2021nzu}.

\subsection{Cost function and observables}

The cost function (${\cal F}$) is needed to optimize several parameters of the neural network in Eq.~\eqref{Eq:FNN}.
The convenient form of the cost function for the POM is
\begin{align}
    {\cal F}[z] &= \int d^{N_{\rm input}} t \, |e^{i\theta(t)}-1|^2 \times |J(t) \, e^{-S(t)}|, 
    \label{eq:cost_function}
\end{align}
where $J(t)$ means the Jacobian of the input and complexified variables and $S$ represents the action.
$\theta(t)$ is the total phase defined by $e^{i \theta(t)} = J(t) e^{-S(t)} / |J(t) e^{-S(t)}|$. 
Minimization of the cost function ($\ref{eq:cost_function}$) provides a modified integral path, which usually makes the Boltzmann weight, $e^{-S}$, being complex.
The phase reweighting is required as,
\begin{align}
    \langle {\cal O} \rangle
    &= \frac{\langle {\cal O} e^{i\theta} \rangle_\mathrm{pq}}
            {\langle e^{i\theta} \rangle_\mathrm{pq}},
    \, \, \,
    \langle {\mathcal O} \rangle_\mathrm{pq}
    := \frac{1}{Z}
       \int \mathcal{D}U \,
         \left[ {\mathcal O} \,
                |J \, e^{-S}|
         \right]_{\mathcal{U} \in \mathcal{C}}
\label{eq:pq}
\end{align}
where ${\cal O}$ represents any operator such as the plaquette.
$\langle \cdots \rangle_\mathrm{pq}$ means the phase-quenched expectation value and $Z$ is the partition function.
The denominator, $\langle e^{i\theta} \rangle_\mathrm{pq}$, is so-called the average phase factor which dominates the statistical error of observables with the phase reweighting.
The absolute value $|J e^{-S}|$ in Eq.\,(\ref{eq:pq}) is used as the Boltzmann weight.
Since $|J e^{-S}|$ is definitely real, we can perform the MC simulation exactly.
Another way is to use $\exp(-\mathrm{Re}\,S)$ as the Boltzmann weight.
In this case, the upper bound of the average phase factor $\langle e^{-(S - \ln J - \mathrm{Re}\,S)} \rangle_\mathrm{R}$ is not necessarily $1$ where $\langle \cdots \rangle_\mathrm{R}$ means the expectation values with the Boltzmann weight, $e^{-\mathrm{Re}\,S}$.

\section{Improvements}
\label{sec:improve}

We explain two improvements to the path optimization: the gauge-covariant neural network to represent the modified integral path, and the approximation of the Jacobian in the learning process.

\subsection{Gauge-covariant neural network}

One of the possible ways to suitably treat the gauge symmetry in the neural network is to use the gauge-covariant neural network~\cite{Tomiya:2021ywc}.
The neural network is represented by the gauge-covariant function using some smearing process.
We employ the following Stout-like smearing,
\begin{align}
    \underbrace{A_\mu(n) \to U_\mu(n)}_{\mathrm{input\,layer}} \to \underbrace{ {\tilde {\cal U}}_\mu^{(1)} (n)\to \cdots}_{\mathrm{hidden\,layer}} \to \underbrace{{\cal U}_\mu(n)}_\mathrm{output\, layer},
\label{eq:gauge_covariant_nn}
\end{align}
where $U_\mu(n)$ denotes the original link variable, ${\tilde{\cal U}}_\mu(n)$ means the smeared (complexified) link variable defined as
\begin{align}
    {\tilde{\cal U}}_\mu^{(l+1)}(n) = \exp[ iW^{(l)}_\mu(n) ]\,{\tilde{\cal U}}_\mu^{(l)}(n),
    \label{eq:smear}
\end{align}
with
\begin{align}
    W^{(l)}_\mu = \sum_{\nu \neq \mu} (\rho^{(l)}_+ {\cal P}^{(l)}_{\mu\nu} + \rho_-^{(l)} {{\cal P}^{(l)}_{\mu\nu}}^{-1}),
    \label{eq:Stout_smearing}
\end{align}
here $l$ means the number of smearing steps in the hidden layer, and $\rho^{(l)}_\pm \in \mathbb{C}$ are parameters optimized via the back-propagation method.
We use the Stout-like smearing twice in the hidden layer to complexify the dynamical variables.
The complexified plaquette ${\cal P}^{(l)}$ in Eq.\,(\ref{eq:Stout_smearing}) is calculated from the link variable of that in the step $l$.
The smearing process combines the link variables again and again as $l$ increases and forms a gauge-covariant neural network.
This network does not correspond to the fully connected layer, but to the residual connection.
The link variables in the previous step is reflected in the total factor in Eq.\,(\ref{eq:smear}), and thus may be robust against the vanishing gradient problem.

Note that the gauge-covariant neural network has a smaller number of parameters than those of the neural network with gauge variant and gauge invariant inputs.
The number of the gauge-covariant neural network parameters is $2 \, l N_{\rm vol}$, while that of the neural network with gauge invariant input is $(N_{\rm input} N_{\rm hidden} + N_{\rm hidden}) + (2 N_{\rm hidden} N_{\rm vol} + 2 N_{\rm vol}) + 2 N_{\rm vol}$, where $N_{\rm input}, N_{\rm hidden}$ are proportional to $N_{\rm vol}$.

\subsection{Approximation of Jacobian}

The calculation cost of the Jacobian in the modification of the path integral contour is rather high, ${\cal O} (N^3)$ for the system with the degrees of freedom $N$.
Reduction of the cost of the Jacobian is necessary for a large scale simulation using the POM.

There are several studies for reducing the Jacobian calculation cost in the MC update:
the Gaussian or real approximations~\cite{Alexandru:2016lsn} that reduce ${\cal O} (N^3)$ to ${\cal O}(N)$,
the implementation of the Grady algorithm in the holomorphic gradient flow~\cite{Alexandru:2017lqr} that reduces ${\cal O} (N^3)$ to ${\cal O}(N^2)$,
the diagonal ansatz of the Jacobian matrix~\cite{Alexandru:2018fqp} that reduces ${\cal O} (N^3)$ to ${\cal O}(N)$,
the nearest neighbor lattice site ansatz~\cite{Bursa:2018ykf} that reduces ${\cal O} (N^3)$ to ${\cal O}(N)$,
the introduction of the worldvolume approach~\cite{Fukuma:2020fez} which requires only a combination of the Jacobian and the vector, i.e., $(J v)$ evaluated using
the anti-holomorphic flow without the explicit form of the Jacobian in the MC update and thus the cost is reduced from ${\cal O} (N^3)$ to $ {\cal O}(N^2)$,
and the application of complex-valued affine coupling layers~\cite{Rodekamp:2022xpf} that reduces ${\cal O} (N^3)$ to ${\cal O}(N)$, in principle.

As a first attempt, we use the simplest treatment for the Jacobian calculation in the path optimization method.
We completely neglect the Jacobian contribution in the learning part.
The numerical cost for the Jacobian calculation is reduced from ${\cal O} (N^3)$ to ${\cal O}(1)$.
It should be noted that we need the Jacobian calculation in the evaluation part of observables.
We will see that this drastic approximation in the path optimization method can work well compared to the full calculation.

\section{Numerical result}
\label{sec:numerical_result}

\begin{figure}[t]
 \centering
 \includegraphics[width=7.5cm]{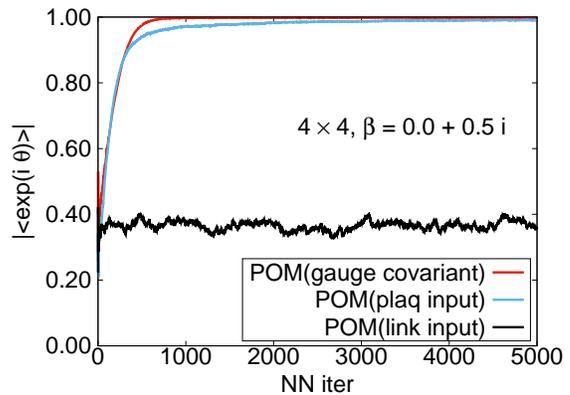}
 \caption{Comparison of the average phase factor by the gauge-covariant neural network with those by the gauge invariant and gauge variant link inputs to the POM~\cite{Namekawa:2021nzu} as functions of the neural network iteration at $\beta = 0.5 i$ on $4 \times 4$ lattice.}
 \label{fig:NN_iter-apf}
\end{figure}

\begin{figure}[t]
 \centering
 \includegraphics[width=7.5cm]{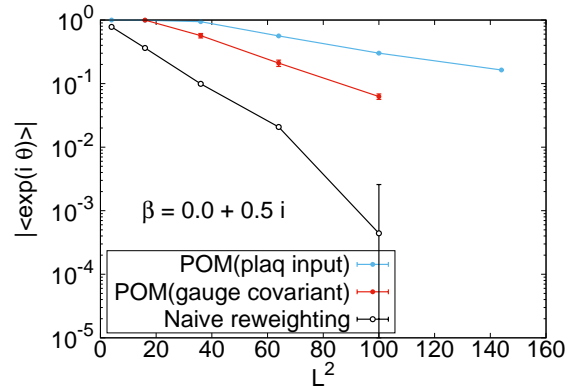}
 \caption{Volume dependence of the average phase factor by the gauge-covariant neural network as well as the gauge invariant input to the POM and the naive reweighting method~\cite{Namekawa:2021nzu} at $\beta = 0.5 i$ on $4 \times 4$ -- $12 \times 12$ lattices.}
 \label{fig:L2-apf}
\end{figure}

\begin{figure}[t]
 \centering
 \includegraphics[width=7.5cm]{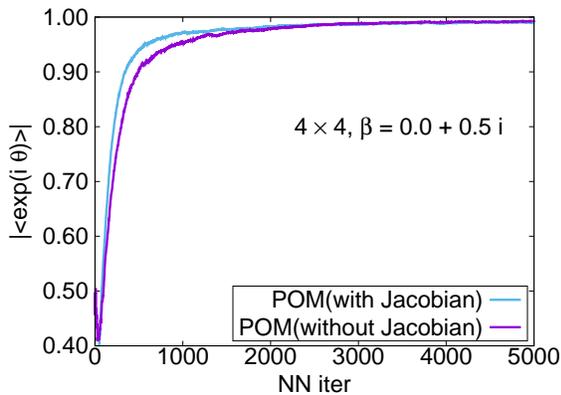}
 \caption{Comparison of the POM using the gauge invariant input with and without Jacobian in the neural network at $\beta = 0.5 i$ on $4 \times 4$ lattice.}
 \label{fig:NN_iter-apf_Jacobian}
\end{figure}

\begin{figure}[t]
 \centering
 \includegraphics[width=7.5cm]{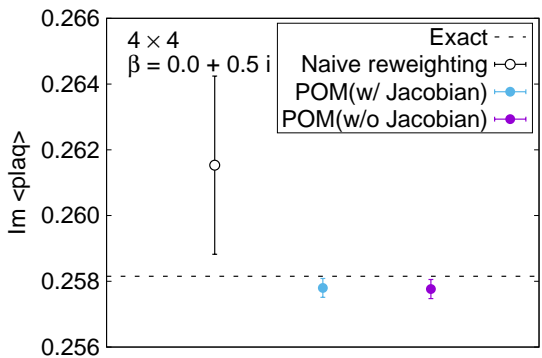}
 \caption{Comparison of the imaginary part of plaquette expectation values by naive reweighting and POM with and without Jacobian in the neural network at $\beta = 0.5 i$ on $4 \times 4$ lattice.}
 \label{fig:imag_plaq_Jacobian}
\end{figure}

We employ $2$-dimensional $\text{U}(1)$ gauge theory with a complex gauge coupling constant as a laboratory to test the improvements for the path optimization method.
The action is
\begin{align}
  S = - \frac{\beta}{2} \sum_n \left( P_{12}(n) + P_{12}(n)^{-1} \right),
  \label{Eq:action}
\end{align}
where $\beta = 1 / g_0^2$.
A complex value of $\beta$ causes a sign problem.
The analytic solution can be found in Refs.~\cite{Wiese:1988qz,Rusakov:1990rs,Bonati:2019ylr}.
The theoretical setup is the same as that in Ref.\,\cite{Namekawa:2021nzu}.
We only change some parts of the POM for the improvements mentioned above.
50000 gauge configurations are generated by the Hybrid MC with its error estimation by the Jackknife procedure.
Adam optimizer~\cite{Kingma:2014vow} is used for the neural network.

Figure\,\ref{fig:NN_iter-apf} shows a comparison of the average phase factor by several inputs to the POM as functions of the neural network iteration.
The POM using the gauge-covariant neural network and gauge-invariant inputs successfully enhances the average phase factor up to 0.99, while that using the link input does not.
It clearly shows the importance of the gauge symmetry in a neural network.
The neural network respecting the gauge symmetry is efficient to control the sign problem by the POM.

The volume dependence of the average phase factor is plotted in Fig.\,\ref{fig:L2-apf}.
Both the gauge-covariant neural network and the gauge invariant input lead to enhancement of the average phase factor and milder volume dependence.
Lower enhancement of the average phase factor by the gauge-covariant neural network may be related to the fact that the gauge-covariant neural network has fewer network parameters than the neural network with the gauge invariant input.

Figure\,\ref{fig:NN_iter-apf_Jacobian} presents a comparison of the average phase factor by the POM using the gauge invariant input with and without the Jacobian calculation in the neural network.
The POM without the Jacobian gives less enhancement of the average phase factor. 
However, the decrease is tiny in our setup.
We also compare the expectation values of the imaginary part of the plaquette in Fig.\,\ref{fig:imag_plaq_Jacobian}.
All results are consistent with the analytic value.
The naive reweighting gives a huge error, whereas the POM with and without the Jacobian gives a much smaller error, indicating that the sign problem is under control.
The difference between errors of the POM with and without the Jacobian is 1\%.
It suggests that the $J = 1$ approximation in the neural network works well.

\section{Summary}
\label{Sec:Summary}

We explored the improvement in the path optimization method for the gauge theory by the following two methods:
\begin{enumerate}
   \item Gauge-covariant neural network.
   \item Approximation of the Jacobian in the learning process.
\end{enumerate}

Using Method 1, the neural network is gauge-covariant by definition.
Due to this property of the network, we do not need the gauge fixing to enhance the average phase factor, unlike Ref.~\cite{Kashiwa:2020brj}.
The enhancement is not as impressive as that of the gauge invariant input~\cite{Namekawa:2021nzu} probably due to the fewer network parameters.
However, the gauge-covariant neural network has an advantage that it can be applied to the non-Abelian theory in a straightforward manner.
Similar performance evaluation in the non-Abelian theory is desirable, especially in QCD at finite density.

Using Method 2, we can significantly reduce the numerical cost.
Neglecting the Jacobian contributions in the learning part reduces the cost from ${\cal O}(N^3)$ to ${\cal O}(1)$.
Even with this drastic approximation, the path optimization can proceed well, and the average phase factor is sufficiently increased.
The drawback is a larger statistical error, but the increase is small, 1\% in our calculation.
We can extend Method 2 by first repeating the learning process without Jacobian contributions and then performing the learning process with the exact Jacobian calculation.
This way has similarity with the pre-training known in the machine learning community and may accelerate the learning of networks.

By using the above two improvements, we can control the gauge symmetry and reduce the numerical cost significantly in the path optimization method at least for the $2$-dimensional $\text{U}(1)$ gauge theory with complex coupling constant.
These improvements may lead us to an exploration of the non-Abelian theory, such as $\text{SU}(2)$ and $\text{SU}(3)$, which has a more serious sign problem and requires a higher numerical cost.
We hope to report this attempt in the future.

\begin{acknowledgments}
Our code is based in part on LTKf90~\cite{Choe:2002pu}.
This work is supported by Japan Society for the Promotion of Science (JSPS) KAKENHI Grant Numbers 
JP19H01898,
JP21H00121,
JP21K03553,
JP22H05112
and by National Natural Science Foundation of China (NSFC) Grant Number 12075061.
\end{acknowledgments}

\appendix
\section{Improvements with link-variable input}
\label{sec:link}

We revisit efficiency of the link-variable input to the neural network.
As already shown in Fig.~\ref{fig:NN_iter-apf}, the average phase factor does not increase with the link-variable input in our setup.
According to the universal approximation theorem~\cite{Cybenko:1989,Hornik:1991251}, however, any continuous function can be expressed in a neural network with a large number of units in hidden layers.
The failure of the optimization implies lack of the expressive power in the present neural network, or lack of enough training data for the network to learn the gauge symmetry.

We examine the improvements in the expression power of the network.
The default network parameters (hyper parameters) are as follows~\cite{Namekawa:2021nzu}.
The batch size (number of configurations used in calculating the derivatives in the stochastic gradient method) is $N_\mathrm{batch}=10$,
the number of the hidden layers is $N_\mathrm{lay}=1$,
and the number of units in the hidden layer is $N_\mathrm{unit}=10$ on a $2\times2$ lattice.
These hyper parameters are increased to explore enhancement of the average phase factor.

We also test enlargement of the gauge configurations by use of the gauge transformation.
After generating gauge configurations in the standard way, we adopt the gauge transformation to the configurations.
It produces additional configurations though equivalent under the gauge symmetry.
This approach is similar to the data augmentation known in the machine learning community.
Through the data augmentation, we can increase the training data and avoid the overtraining. The neural network with the enlarged gauge-transformed configurations is expected to learn the symmetry of the input data better.
With this expectation, we have performed the optimization using the enlarged configurations.

\begin{figure}[t]
 \centering
 \includegraphics[width=7.5cm]{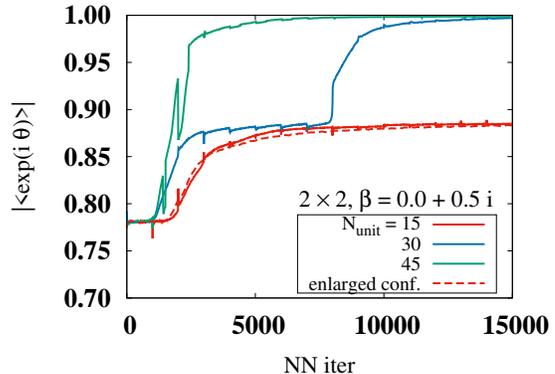}
 \caption{The average phase factor by the POM with the gauge link input as a function of the neural network iteration at $\beta = 0.5 i$ on a $2 \times 2$ lattice. The number of units in the hidden layer is $N_{\rm unit} = 15,30,45$.
 The result of the enlarged configurations using $N_{\rm unit} = 15$ is also plotted.}
 \label{fig:NN_iter-apf_link_input}
\end{figure}

In Fig. \ref{fig:NN_iter-apf_link_input}, we show the average phase factor using the link input with increased hyper parameters $N_\mathrm{batch}=1000$, $N_\mathrm{lay}=5$, and $N_\mathrm{unit}=15,30,45$ on a $2\times2$ lattice.
The enhancement of the average phase factor with larger $N_\mathrm{batch}$, $N_\mathrm{lay}$ and $N_\mathrm{unit}$ is observed, as expected from the universal approximation theorem.
We also find the gauge configurations enlarged by a factor of 100 using the gauge transformation give no additional enhancement, and hence less effective.
With $N_\mathrm{unit}=30$ and $45$, the average phase factor goes through a sudden increase once or twice and reaches $0.99$ after 12000 and 5000 iterations, respectively.
The $N_\mathrm{unit}=15$ case seems to require a larger number of the neural network iterations for the average phase factor of 0.99.

From the numerical results, we can clearly see that even a simple neural network using the link input can learn the gauge symmetry, though the cost of the optimization is significantly higher than that of the gauge-invariant input and the gauge-covariant neural network.

\bibliography{ref.bib}

\end{document}